\documentclass[12pt]{iopart}
\usepackage{iopams}  
\begin{document}
 \def\cF{{\cal F}} \def\cG{{\cal G}}
  \def\cM{{\cal M}}
\def\IR{\relax{\rm I\kern-.18em R}}
\def\half{{1 \over 2}}
\def\um{{\underline{m}}}
\def\un{{\underline{n}}}
\def\up{{\underline{p}}}
\def\uq{{\underline{q}}}
\def\ui{{\underline{i}}}
\def\uj{{\underline{j}}}
\def\uk{{\underline{k}}}
\def\ul{{\underline{\ell}}}
\def\V{{V}}
\def\T{T}

\def\deom{{\Delta\Omega}}
\def\cv{\mathcal{V}}

\def\id{\relax{\rm 1\kern-.35em 1}}

\def\eq#1{(\ref{#1})}  
\newcommand{\cV}{\mathcal{V}}

\newcommand{\fsl}{\mathfrak{sl}}  
\newcommand{\fgl}{\mathfrak{gl}}  
\newcommand{\fe}{\mathfrak{e}}  
\newcommand{\fo}{\mathfrak{o}}  
\def\R{\mathbb{R}}  
\def\C{\mathbb{C}}  
\def\Z{\mathbb{Z}}  
\def\Hb{\mathbb{H}}  
\def\cM{\mathcal{M}}

\title{Gauge Charges from Supergravity}

\author{L. Andrianopoli}

\address{Centro E. Fermi, Rome, Italy; \\PH-TH-Unit, CERN, CH-1211 Gen\`eve 23, Switzerland and\\   Dip. Fisica, Politecnico di Torino,   C.so Duca degli Abruzzi,  24 I-10129 Torino, Italy \\and
INFN, Sezione di Torino, Italy}
\ead{Laura.Andrianopoli@cern.ch}
\begin{abstract}
Some recent results in the study of four dimensional supergravity flux compactifications are reviewed, discussing in particular the role of torsion on the compactification manifold in generating gauge charges for the effective four dimensional theories.
\end{abstract}

%Uncomment for PACS numbers title message
%\pacs{00.00, 20.00, 42.10}
% Keywords required only for MST, PB, PMB, PM, JOA, JOB? 
%\vspace{2pc}
%\noindent{\it Keywords}: Article preparation, IOP journals
% Uncomment for Submitted to journal title message
%\submitto{\JPA}
% Comment out if separate title page not required
%\maketitle

\section{Introduction}
Many efforts in theoretical physics have been devoted, in the last decades,  to look for a quantum theory describing in a consistent  and  unified way gauge interactions with gravity.
Such a theory, if any, should incorporate the Standard Model of elementary particles but  is also expected to give answers to some problems left open after the huge theoretical and phenomenological successes of the Standard Model, like the understanding of the hierarchy problem, of confinement in QCD, of the tiny value of the cosmological constant, or of  the black-hole entropy.

 In this spirit, superstring theory  appears nowadays as a very promising candidate for the quantum theory underlying Nature:  it is indeed a consistent and finite theory for  quantum gravity,  giving unification of  gauge interactions  with  gravity. It is described by a two dimensional conformal theory defined at the Planck scale, spanning a curved ten dimensional target space. When the spectrum is extended to include solitonic configurations (D-branes \cite{pol}) it may reveal one extra eleventh dimension (M-theory) \cite{m-theory,dual}. 

The massless excitations of superstrings are described, in a low energy limit,
by effective supersymmetric field theories, named supergravity theories, which   contain supersymmetric couplings of gravity with matter and gauge fields.
Supergravity theories may be defined in any space-time dimensions up to eleven, however they get predictive power when interpreted as effective theories for superstrings or M-theory.
 If superstring theory has to provide a natural explanation of interactions
 in our real world, some trick must be present so that
 in particular the theory, at low energies, looks four dimensional.
 The simplest way to do so is to imagine that the vacuum configuration
 for space-time is not ten dimensional Minkowski space, but is instead
 of the form $\IR^{(1,3)}\times M_6$, where $M_6$ is some six-dimensional
 compact manifold, whose size is so small that at the length-scales
 experimented in our low energy world it cannot be observed.
 Then, any tensor field of the ten dimensional theory looks, in the four
 dimensional compactified theory, as a collection of
 fields of various masses and spins \cite{cj} (the procedure of compactification from
 higher dimensions to four was first studied in five dimensions by Kaluza
 and Klein \cite{kakl}).
 In particular, the ten dimensional fields give rise to a
 set of massless
 four dimensional scalars ({\it moduli})  describing metric and topological
 properties of the compactification manifold $M_6$.
 Any choice of the vacuum background of string theory
 selects one among a lot of possible inequivalent compactified space-times,
 and this implies giving a particular expectation value to the scalars.
 Note that these scalars are just the ones
 appearing in supergravity theories.
 Giving them some particular vacuum expectation values means, going down to very low
 energies, fixing the values of the  physical observables in
 the Standard Model describing strong and electro-weak interactions.

   We are going to focus on this intertwining role of supergravity
    between the stringy (Planck) scale 
   and  low energy   physics.
   We will discuss in particular how gauge interactions arise in supergravity theories, with particular attention to the role of the torsion in the compactification manifold as gauge charge.

This contribution is organized as follows: In section 2 the role of scalar fields and the way gauge charges appear in supergravity are surveyed; in section 3 the example of no-scale models as gauged supergravity theories is discussed, with particular attention to the Scherk--Schwarz model,  while  section 4 focuses on the role of torsion on the compactification manifold in flux compactifications and on its relation to the Scherk--Schwarz model. Conclusions and outlook are left to the concluding section.
The bibliography section is not exhaustive; somewhere we preferred to give reference to review papers where more extended references on the subject may be found.  For a more complete list of references we refer to \cite{noi}.

%%%%%%%%%%%%%%%%%
 %%%% section 2    %%%%%
 %%%%%%%%%%%%%%%%%

\section{Scalars in supergravity and gauging}

A   peculiar feature of supergravity is that the  spectrum includes
 fermionic  and   bosonic  
fields with
  spin $\leq$ 2, among which   scalars. 
 Supersymmetry strongly constrains  the couplings, and in particular it implies that  the  scalars belong to  non linear $\sigma$-models: 
\begin{equation}
\mathcal {L}_{kin|scalars}= g_{\ell m}(\phi) \partial_\mu \phi^\ell  \partial^\mu \phi^m.
\end{equation}
 The scalars play an important role in the field theory,  as they generalize
 the Higgs field of the Standard Model. 
 The low energy observables
(gauge couplings,  mass parameters,  the cosmological constant, etc...) 
are  given in terms of  vacuum expectation values  (v.e.v.) of scalars.
The cosmological constant $\Lambda$, for example, is the  v.e.v. of the scalar potential
\begin{equation}
\Lambda =  < V(\phi)>,\end{equation}
while the gauge couplings and $\theta$-term appear in the lagrangian as the v.e.v. of scalar dependent functions:
\begin{equation}
{\cal L}_{kin|vec}=
  \gamma_{\Lambda\Sigma}(\Phi)
 \cF^\Lambda \wedge {^*}
 \cF^{\Sigma } + 
\theta_{\Lambda\Sigma}(\Phi)  \cF^\Lambda\wedge
\cF^{\Sigma }.
\end{equation}
However, the v.e.v. of the scalar fields fixing the value of the couplings in  the effective supergravity theories are in fact determined by the  choice of a vacuum in the corresponding underlying superstring theory. 
What happened, with the introduction of  superstring theory as a unifying theory
 describing in particular particle physics, is that  the Standard Model problem of having
 many free parameters to be fixed 
 has turned into the problem of an enormous degeneracy of vacua for string
 theory.
 As we can see, in this way we do not gain any predictive power on the
 quantities characterizing our world; however, we obtain the important
 conceptual achievement of unifying gravity with the other interactions
 and to give a natural understanding for the origin of all the parameters
 involved (completely arbitrary in the Standard Model).
 Moreover, there is the hope that some non-perturbative stringy effect,
 inducing supersymmetry breaking, could also lift the vacuum degeneracy
 leaving fewer accessible vacua.

One is then driven to the problem of understanding how gauge charges and interactions arise in  supergravity.
 
  Supergravity theories are built from  symmetry  requirements:  local invariance under the supersymmetry algebra. The bosonic sector includes  gravity,   abelian  gauge fields  and  uncharged   scalars; to get a theory invariant under the local supersymmetry algebra there is no need to introduce  gauge charges.
The gauge charges are introduced  via a  deformation  of the theory named  {\it the
gauging procedure}, which corresponds to promote 
a      global symmetry  of the theory (an    isometry of the scalar manifold) to
  gauge symmetry and to turn on 
interactions with the corresponding gauge fields of the theory (which may become non-abelian).
Consistency of the deformation with  supersymmetry   implies the   presence of a 
 scalar potential, where the
 dynamical information on the model  are codified.

However, if we want to study supergravity theories as effective theories of superstrings, it must be possible to interpret the gauge charges of a given supergravity model  in terms of data corresponding to some superstring background.
For example,  {\em ungauged} supergravity models (with no gauge  charges) appear as the low energy limit of perturbative Type II superstrings compactified on tori without any non-trivial flux for the tensor fields (p-forms) of the theory in the internal directions.
On the other hand,  starting from a ten dimensional theory,  three different mechanisms have been identified to generate gauge charges in the four dimensional effective  supergravity models \cite{km,dst,dkpz,noi}:
\begin{itemize}
\item  by turning on charges corresponding to  non abelian gauge fields  already in the
 ten dimensional  theory (from gauge fields living on D-branes, for Type I or II theories,
  or  from the heterotic sector, for the case of the heterotic string);
\item by turning on  non trivial  p-form fluxes   in the internal compact manifold;
\item 
by turning on charges associated to the   geometry  of the compactification manifold (\`a la Scherk and Schwarz).
\end{itemize}
The emerging scenario is that the first class of charges should contain the gauge group of the Standard Model, while the other two classes of charges should be responsible for supersymmetry breaking, via the super-Higgs mechanism.

 Note that the dualities among string theories still map one into the other the theories in the presence of charges, provided that also the charges are transformed accordingly.

 %%%%%%%%%%%%%%%%%%%%%%
 %%%%%%%%%%%%%%%%%%%%%%
\section{An explicit example: the no-scale models
}
No-scale models are a particular class of supergravity theories emerged in the 80's  for their
 phenomenological interest \cite{bfs,ln}. They have  seen a revived interest from theoretical physicists after that some recent progress in string theory   (D-branes \cite{pol}, Randall--Sundrum scenarios \cite{rs})   allowed to find them a nice interpretation  in terms of the  classes of ten dimensional charges  introduced in the previous section.

No-scale supergravities are models with an automatically vanishing cosmological constant  
(at tree level), and where  supersymmetry is spontaneously broken.  The gravitino mass  $m_{ {3/2}} $  is a sliding scale   at tree level, dynamically fixed by radiative corrections, with possibility of a  hierarchical suppression  with respect to  the Planck scale.
The characteristic features of the various no-scale models are codified in the choice of the gauge group and in the structure of the scalar potential. However, they all share the feature of having a semi-positive definite scalar potential  with flat directions. 
From a purely supergravity point of view, no-scale models
are typically related to gaugings of suitable non-semisimple  
global symmetry groups of the Lagrangian (\emph{flat gaugings})  
\cite{ssgauging0,ssgauging1}. 
A peculiar feature of this kind of models is that the
 spectrum appears separated in an  {\em observable}   sector,
(where the standard Model is defined),  and  an {\em
hidden}   sector (which is decoupled for energy scales   $E<< M_P$  ) interacting with the observable sector only through gravity.

Phenomenologically viable no-scale models have been constructed in  string compactifications \cite{flux1} where the   SM/GUT  lives on space-time filling  D-branes (observable sector)  while the  hidden sector   is given by the   bulk  
 of the  ten dimensional ambient space.
A   hierarchical   supersymmetry breaking \cite{gkp}  with stabilization of (some of the) moduli may be achieved in the bulk (hidden sector)
 by turning on  background fluxes in the compactification manifold.
 Interesting models  of flux  
compactifications have been found mostly in Type IIB theory by switching on  
appropriate RR and NS three-form fluxes in the internal directions  
\cite{flux2}  (other examples were obtained in  heterotic M-theory, due to a non-vanishing G-flux sourced by
the boundaries \cite{krause},  and on the  
Type IIA  front \cite{dkpz,IIAflux}).

Alternatively, no-scale models may be found as well  from  
generalized dimensional reduction \`a la Scherk--Schwarz (SS)  
\cite{ss} of eleven dimensional supergravity  
(or any truncation thereof). This is a generalized type of  
dimensional reduction on tori,
corresponding to a  twist in the boundary conditions of the fields in the internal directions, with a phase dependent on a  global symmetry   of the theory  (SS phase).
The twist originates a  no-scale model  with non abelian couplings  
and a positive definite scalar potential, which, in general,  
yields spontaneous supersymmetry breaking. 
The charges obtained this way appear to be associated to the internal manifold geometry. However until recently they could not find a direct interpretation in the context of flux compactifications of superstring or M theory.  We are going to see in the following  section that such an interpretation is in fact possible in the presence of  torsion  in the internal manifold.
Let us first briefly review what the SS mechanism is.

%%%%%%%%%%%%%%%%%%%%%
%%%%%%%%%%%%%%%%%%%%%

\subsection{The Scherk-Schwarz mechanism}\label{ss1}
To describe how the mechanism works, let us consider  the simple example  of a free complex scalar field $\phi(x,y)$ on  $\IR^{1,3}\times S^1$, with lagrangian 
\begin{equation}
{\mathcal L}= \half \partial _{\hat \mu}\phi \partial ^{\hat \mu}\phi^* ,
\label{lagr}
\end{equation}
where
$ \hat\mu = (\mu,4);\, \mu = 0,\dots ,3;\, x^4=y$.
 We want to interpret this model from a four dimensional point of view, supposing that the $S^1$ has a tiny radius. However, instead of the Kaluza--Klein (KK) ansatz, let us impose the following generalized ansatz of dimensional reduction:
 \begin{equation}
\phi(x,y) = {\rm e}^{{\rm i} my} \sum_{n=-\infty}^{+\infty}
\phi_n(x) {\rm e}^{{\rm i} ny/ R}\, , \quad (y \sim y+ 2\pi R). \label{ans}
\end{equation}
We observe that  this ansatz is not well defined, because $\phi(x,y)$ is multivalued on $S^1$. However, since the lagrangian \eq{lagr} is invariant  under  global  $U(1)$ transformations
\begin{equation}
 \phi \to {\rm e}^{{\rm i}\alpha}\phi
 \end{equation}
 then the   phase 
  $  {\rm e}^{{\rm i}2\pi mR}$  has no physical effects.
The  five dimensional kinetic lagrangian \eq{lagr}  generates, in the dimensional reduction to four dimensions, a  ``scalar potential'' (a mass term, in fact, in this case)
 $V=m^2 >0$.
 All the masses are shifted by $ m $, and the  zero-mode has a mass $m$.
 
The key ingredient, which allows to implement the SS mechanism and to introduce in the dimensionally reduced theory the ``charge'' $m$, is the global invariance of the lagrangian \eq{lagr} under the symmetry group $U(1)$. The charge is then introduced as a $y$-dependent phase in the dimensional reduction ansatz \eq{ans}. The same mechanism, applied to theories invariant  under more general  global symmetries, allows to turn on more general charges related to the internal geometry.

In particular, let us now study the case of generalized dimensional reduction of $4+n$ dimensional  gravity on 
 $\cM_{4+n} =\IR^{1,3}\times
S^1\times T^{n-1}$. The result will be compared, in the next section, with an ordinary KK dimensional reduction  of $4+n$ dimensional  gravity on the same manifold 
 $\cM_{4+n} $ but in the presence of a torsion background.

 Let us  suppose to formally  perform the reduction in two steps:
\begin{itemize}
\item first consider the    KK   compactification    to {\em five dimensions}    on
 $T^{n-1}$.    One gets a five dimensional  theory on $\IR^{1,3}\times
S^1$ with   global invariance
 $SL(n-1)$, which is the   isometry group of the torus $T^{n-1}$.
 \item
 then we may use this global invariance  for building a   SS phase-matrix  for
  a {\em generalized}  
dimensional reduction {\em from   five to four dimensions}.
\end{itemize}

From the $4+n$-dimensional vielbein one gets the following  fields in the theory reduced to five dimensions:
the five dimensional vielbein $    V_{\hat \mu}{}^{\hat a}$, $n-1$ vectors $    B_{\hat
\mu}^{{m}}$  and $\frac{n(n-1)}2$ scalars $    \phi^i ,     \sigma$   parametrizing the coset $\frac{SL(n-1)}{O(n-1)}\times
O(1,1) $, whose coset-representatives are the internal components of the $4+n$ dimensional vielbein,   $V_{m}{}^i (\phi^i,\sigma) $.

  The SS ansatz for the dimensional reduction (restricted to the zero-modes)
from five to four dimensions is (we take $x^4 \equiv y^1$ and no dependence of the fields on the extra internal coordinates $x^5=y^2,\cdots ,x^{n-1}=y^{n-4}$)
\begin{eqnarray}
 %\begin{cases} 
 B_{\hat \mu}^{m}(x,y^1)
& = & U^m_{\ n}(y^1)B_{\hat \mu}^n (x) \nonumber\\
V_{m}{}^i(x,y^1) &=& (U^{-1})_m^{\ n}(y^1) V_{n}{}^i(x)
%\end{cases}
\label{ssansatz}
\end{eqnarray}
 where  $U=\exp[{My^1}]$,  $M \in SO(n-1)$.
The theory reduced to  four dimensions has  non abelian gauge charges  which come
from the dimensional reduction of the  Levi--Civita connection  on $\cM_{4+n}$, using  the generalized ansatz \eq{ssansatz}. In
particular one finds \cite{ss,ssgauging0} ($\um =(4,m)$; $\ui =(4,i)$)
\begin{eqnarray}
 F^\um_{\mu\nu} &=& \partial_{[\mu} B_{\nu]}^{{\um}}+ 
f^\um_{\ \un\up}  B_{[\mu}^{{\un}}B_{\nu]}^{{\up}}\\
D_\mu V_{{\um}}{}^\ui &=& \partial_\mu  V_{{\um}}{}^\ui +  f^\un_{\
\um\up}  B_{\mu}^{{\up}}V_{{\un}}{}^\ui
\end{eqnarray}
 with structure constants given by the    SS phases:  
 \begin{equation}
 f^m_{\ 4n}= M^m_{\
n} .
\end{equation}

Let us make an observation which will be important in the next section. If the theory  in $4+n$
dimensions contains also  p-forms besides the metric, then the SS prescription requires that  the scalars coming from the dimensional reduction of  the p-forms have to be rotated as well with the SS phases.
\\
For the case, {\it e.g.},   of a one-form $A_m$ one has
 \begin{equation}
A_m(x,y^1) =  (U^{-1})_m^{\ n}(y^1)  A_n(x) ,
\end{equation}
so that
\begin{equation}
F_{4m}(x,y^1) =  M_m^{\ n} A_n(x) . \label{fss}
\end{equation}

We may then conclude that the SS  mechanism is surprisingly efficient in introducing geometrical charges in the theory. However it appears as a bit artificial prescription, which is  not completely satisfying from the  microscopical point of view.

 We are going to see in the next section that the SS mechanism may be understood as an ordinary flux compactification if we assume the presence of a constant background  torsion  in the internal manifold.
 
 %%%%%%%%%%%%%%%
 %%%%%%%%%%%%%%%
  \section{The SS mechanism as a flux compactification with internal torsion}

The purpose of the present section is to review how theories  
originating from  SS dimensional reduction  may alternatively be  
found as the result of a KK dimensional reduction in  
the presence of an internal torsion. In particular we shall consider a $D+n$  
dimensional pure gravity theory compactified to $D$ dimensions on  
a $n$--torus $T^n$ showing  
 that, by switching on an appropriate constant  
background torsion in the torus  (a {\em torsion  
flux}), we may generate the same $D$-dimensional theory as the one described in section 3.1,
originating  
from a SS reduction from  $D+1$ dimensions with global  
symmetry generator chosen within the global symmetry algebra  
$\fsl(n-1,\R)$. This is based on \cite{noi}.

The relevance of torsion in flux compactifications was recently understood  in the study of  T-duality  between Type IIB and Type IIA  superstrings  \cite{louis,kachru}. Indeed it was found that the  T-dual of the charge associated to the flux of  $H^{(3)}_{NS}$ is not a p-form flux but internal torsion.
Our  geometrical interpretation allows  to study also the no-scale  SS models  in the context of flux compactifications of superstring (or M) theory.

 A delicate point in our analysis is the notion of gauge invariance in the presence of torsion. Indeed,   we found that to have a precise identification of the SS models in terms of torsion a   generalization of the notion of gauge transformation   is needed, corresponding to the  coupling of the torsion to the gauge fields.
 
\subsection{Coupling  torsion - gauge fields: the general idea}
The torsion tensor may be defined as an antisymmetric contribution 
\begin{equation}
T_{\;\;MN}^P\equiv \tilde\Gamma_{[MN]}^P=\frac 12 (\tilde\Gamma_{MN}^P-\tilde\Gamma_{NM}^P)
\end{equation}
 to the affine connection $\tilde\Gamma$. 
 In this case,  the request of a metric  
connection ($\nabla_P\,g_{MN}=0$) determines the  
connection coefficients as \begin{equation}
\tilde\Gamma_{MN}^P=\Gamma_{MN}^P+K^P_{\phantom{P} MN},
\label{connection}
\end{equation}
 where $\Gamma_{MN}^P =\Gamma_{NM}^P $ are the  
coefficients of the Levi-Civita connection while
\begin{equation}
K^P_{\phantom{P} MN}=\frac 12(T_{\phantom{P}  
MN}^P-T _{M\phantom{P} N}^{\phantom{P} P}-T_{N\phantom{P}  
M}^{\phantom{P} P})
\end{equation}
 contains the torsion contribution. For further definitions and conventions, we refer to \cite{noi}.
  
To discuss the origin of the coupling of the torsion to gauge fields let us consider, as an example, the case of a vector field $A_M$.  
The  field strength feels the effect of the torsion background via  
the principle of general covariance:  
\begin{equation}  
F_{MN}  
\equiv \nabla_{[M}A_{N]}=  
\partial_{[M}A_{N]}+T^P_{\ MN}A_P.  
\label{torsionfull}  
\end{equation}  
When $T\neq 0$, $F$ is not invariant  
under the usual gauge transformation
\begin{equation}  
\delta A_M(X) = \partial_N \Lambda(X).  
\end{equation}  
However, it is possible \cite{hrrs} to make  the torsion compatible with the  
presence of gauge fields by  introducing the generalized definition  
of gauge transformation  
\begin{equation}  
\delta_C A_M (X)= C_M^{\ N}(X) \partial_N \Lambda(X)  ,
\label{generalgaugetransf}  
\end{equation}  
where the point-dependent matrix $C_M^{\ N}(X)$ has to be  
constrained by the request of (generalized) gauge invariance of the field strength  
\begin{equation}  
\delta_C F_{MN}=0.  
\end{equation}  
The general solution, found in \cite{hrrs} for a  
gauge field in $d$ dimensions coupled to torsion in $d$ dimensions  
is
\begin{equation}C_M^{\ N} =\delta_M^N e^\phi \,, \qquad T^P_{\ MN}= \delta^P_{[M} \partial_{N]} \phi.  
\label{stringenttorsion}\end{equation}
There  
is only one torsion degree of freedom allowed, the scalar field  
$\phi$.
Let us now see  that 
this restriction may be largely relaxed if the  idea of \cite{hrrs} is combined with the  Kaluza--Klein ansatz in a dimensional reduction context,  and that it is then possible to turn on an internal torsion with many independent components. The different components of the torsion  generate geometrical charges that reproduce the   SS phases  while giving them a   geometrical interpretation .
 %%%%%%%%%%%%%%%%%
 \subsection{Coupling torsion / gauge fields and dimensional reduction} 

Let us consider  a gauge field $A_M$  in the $4+n$ dimensional manifold 
$\cM_{4+n}=\IR^{1,3}\times T^n$ = $\IR^{1,3}\times S^1\times
T^{n-1}$, with coordinates $X^M= (x^\mu,y,y^m)$,
 $m=1,\cdots n-1  \in T^{n-1}$ and $ \mu = 0,1,2,3 \in \IR^{1,3}$; y parametrizes $S^1$ (it is the same manifold as in section 3.1).
 We ask that the
  KK ansatz holds true:
\begin{equation}
A_M =A_M (x)\,; \quad \Lambda = \Lambda(x) .
\end{equation}
 We suppose to have a constant torsion  background on the torus $T^n$, and that the gauge field is coupled to the torsion as in \eq{generalgaugetransf}. 
Then,
  $F=\nabla A$ is not gauge invariant, but it is invariant under the  
  generalized gauge transformation  \`a la HRRS :
\begin{equation}
\delta_C A_M =  C_M^{\ N} \partial_N \Lambda =
  C^{\phantom{M}  \nu}_{  M}  \partial_{\nu}  \Lambda .
\end{equation} 
If we further assume:
\begin{equation}
C_M^{\ N}= C_M^{\ N}( y ),
\end{equation} 
 then the generalized gauge invariance condition  $\delta_C F =0$
has now more general solutions (see \cite{noi} for details), allowing the following non-zero entries for the matrices $C$ and $T$:
\begin{equation}
C_M^{\ N} \,:\, (C_\mu^{\ N}, C_M^{\ n})\,; \quad
T^M_{\ NP}\,:\, (T^\mu_{\ \nu 4},  T^m_{\ NP})
\end{equation} 
 with the only constraint
\begin{equation}
C_\mu^{\ \nu}(y)=
\delta_\mu^{ \nu} e^{\phi(y)} \,; \quad T^\mu_{\ \nu 4}=\frac 12
\delta^\mu_{\nu}\partial_{4  }\phi(y).
\end{equation}

By choosing
 \begin{equation}T^m_{\ 4n}=M^m_{\ n} \in SO(n-1) , \quad (\phi=0),
 \label{tb}
 \end{equation}
we have a consistent theory, where
\begin{equation}
 \delta_C A_\mu = \delta A_\mu =\partial_\mu \Lambda \,, \quad \delta_C A_4 =0\,, \quad \delta_C A_m =0.
\end{equation}
Moeover it gives
\begin{equation} 
F_{4n}=M^m_{\ n} A_m  
\end{equation}
which precisely reproduces the effect of the geometrical SS phase for a vector, equation 
\eq{fss}.

Let us now consider the KK dimensional reduction of  the metric in the presence of the  the
  torsion background \eq{tb}, and let us compare the result with the one  of Section \ref{ss1}.
  As in section \ref{ss1}, the theory reduced to  four dimensions has  non abelian gauge charges  which come
from the dimensional reduction of the affine connection  on $\cM_{4+n}$, and we find the same result as before
\begin{eqnarray}
 F^\um_{\mu\nu} &=& \partial_{[\mu} B_{\nu]}^{{\um}}+ 
f^\um_{\ \un\up}  B_{[\mu}^{{\un}}B_{\nu]}^{{\up}}\\
D_\mu V_{{\um}}{}^\ui &=& \partial_\mu  V_{{\um}}{}^\ui +  f^\un_{\
\um\up}  B_{\mu}^{{\up}}V_{{\un}}{}^\ui
\end{eqnarray}
However, the interpretation is now different. In this case the charges do not come from the dimensional reduction of the metric (which is in fact an ordinary KK reduction) but from the torsion contribution to the affine connection \eq{connection}, which, differently from section \ref{ss1}, in this case is  not any more Levi--Civita. 
 The structure constants of the four dimensional gauge theory  are now given by the components of the torsion  
 \begin{equation}f^m_{\ 4n}= T^m_{\ 4n}.
 \end{equation}
The comparison between the two interpretations is summarized in the  table below.
$$
\hskip -1.5cm\begin{array}{c||c|c}
& \mbox{S-S}   & \mbox{  Torsion } \cr \hline
 f^m_{\ 4n}  =& M^m_{\
n}=\mbox{SS phases} & T^m_{\ 4
n}=\mbox{torsion}
\cr
 F_{4m}  :& A_m(x,y) = (e^{-M y})^m_{\ n}A_n(x) & F_{4m}=  T^m_{\ 4
n}A_n \cr
 \V_m^{\ i}  : & (e^{-M y})^n_{\ m}V_n^{\ i}(x) & V_n^{\ i}(x) \cr
 \Gamma^M_{\ NP}  : & \Gamma^M_{\ [NP]}=0 &  \Gamma^M_{\ [NP]}\neq 0
\end{array}
$$
\section{Conclusions}

We discussed how   gauge charges and interactions   do emerge in supergravity.
We focused in particular on the role of the   charges related to the geometry, and shown that  the SS  mechanism is in fact an ordinary compactification in the presence of an internal torsion background.
The scenario that seems to emerge is that all the bulk charges, responsible for supersymmetry breaking, may be generated by   compactifications in the presence of  fluxes and  internal torsion.
 
 There are various directions for future investigations,  
which include a similar analysis of theories corresponding to  
compactifications on more general manifolds,  
the construction of a $D=4$ gauged supergravity deriving from a  
general torsion background, and  the effects of a dynamical  
torsion on this setting.  
 
% BibTeX users please use
% \bibliographystyle{}
% \bibliography{}
%
% Non-BibTeX users please use

\end{document}